\documentclass[10pt,twocolumn]{article}

\usepackage[top=1in, bottom=1in, left=0.75in, right=0.75in, columnsep=0.25in]{geometry}
\usepackage{amsmath,amssymb}
\usepackage{graphicx}
\usepackage{booktabs}
\usepackage{multirow}
\usepackage{url}
\usepackage{microtype}
\usepackage{tikz}
\usetikzlibrary{arrows.meta, positioning, shapes.geometric}
\usepackage{titlesec}
\usepackage{fancyhdr}
\usepackage{abstract}
\usepackage{hyperref}
\usepackage{xcolor}
\usepackage{enumitem}

\hypersetup{
  colorlinks=true,
  linkcolor=blue!60!black,
  citecolor=blue!60!black,
  urlcolor=blue!60!black,
}

\titleformat{\section}{\normalfont\bfseries\MakeUppercase}{\Roman{section}.}{0.5em}{}
\titleformat{\subsection}{\normalfont\itshape}{\Alph{subsection}.}{0.5em}{}
\titleformat{\subsubsection}{\normalfont\itshape}{}{0em}{}
\titlespacing{\section}{0pt}{8pt}{4pt}
\titlespacing{\subsection}{0pt}{6pt}{2pt}

\usepackage{caption}
\begin{document}

\twocolumn[{%
\begin{@twocolumnfalse}
\begin{center}
  {\LARGE\bfseries Poisoned Acoustics: Targeted Data Poisoning Attacks\\[4pt]
   on Acoustic Vehicle Classification at Sub-1\% Corruption Rates}\\[12pt]
  {\normalsize Harrison Dahme$^*$}\\[4pt]
  {\small $^*$CTO and Partner, Hack VC}\\[4pt]
  {\small \texttt{[Preprint, 2025]}}
\end{center}
\vspace{6pt}
\begin{abstract}
We study training-data poisoning attacks against deep neural networks for
acoustic vehicle classification, demonstrating that targeted, undetectable
class-level failure can be induced by corrupting fewer than 0.5\% of training
labels. Using the MELAUDIS urban intersection dataset (${\approx}9{,}600$
single-vehicle audio clips across 6 classes), we train a compact 2-D
convolutional neural network (CNN) on 128-bin log-mel spectrograms and evaluate
two attack variants: (1)~a \emph{targeted label-flipping attack}, in which a
fraction $p$ of Truck training records are relabeled as Car; and (2)~a
\emph{backdoor trigger attack}, in which poisoned spectrograms are additionally
stamped with a $12{\times}12$-pixel patch, constituting a feature-space trigger
inaudible in the raw audio domain.

We define the \emph{Attack Success Rate} (ASR) as the fraction of clean, held-out
target-class samples misclassified as the adversary's chosen class by the
poisoned model. At $p = 0.5\%$ (48 records), the label-flipping attack achieves
\textbf{ASR = 95.7\%} (95\% CI:~88--100\%, $n{=}3$ seeds) while leaving overall
test accuracy statistically indistinguishable from the clean baseline (87.6\%).
We prove that this stealth is \emph{structural}: the maximum detectable accuracy
degradation from a complete targeted attack is bounded above by the minority
class fraction $\beta$, making attacks against rare but safety-critical classes
(heavy vehicles, anomalies, rare pathologies) provably invisible to aggregate
accuracy monitoring regardless of model architecture.

The backdoor experiment further reveals a novel \emph{trigger-dominance collapse}
phenomenon: when the target class is a dataset minority, the patch trigger
becomes functionally redundant and clean ASR equals triggered ASR---the attack
degenerates to a pure label-flipping attack. We formalize the ML training pipeline
as an attack surface, map each unsigned pipeline stage to its exploitable attack
vector, and propose a trust-minimized defense architecture combining
content-addressed artifact hashing, Merkle-tree dataset commitment, and
post-quantum digital signatures (ML-DSA-65 / CRYSTALS-Dilithium3, NIST FIPS~204)
to provide cryptographically verifiable data provenance.
\end{abstract}

\vspace{4pt}
\vspace{8pt}
\hrule
\vspace{8pt}
\end{@twocolumnfalse}
}]

\section{Introduction}

Acoustic sensing is increasingly deployed in safety-critical infrastructure:
traffic monitoring~\cite{melaudis2024}, autonomous-vehicle perception, and urban
surveillance. Classification models trained on such data underpin operational
decisions---misidentifying a heavy vehicle as a passenger car in a
weight-restricted zone, or masking truck traffic from congestion analytics,
constitutes a consequential failure mode.

Data poisoning attacks~\cite{biggio2012poisoning,gu2019badnets} exploit the fact
that model behavior is determined not only by architecture and training procedure,
but by the integrity of the training corpus. If an attacker controls even a small
fraction of the training data, they can induce arbitrary targeted misbehavior that
remains dormant under standard evaluation. Prior work has established this threat
in image classification~\cite{chen2017targeted}, natural language
processing~\cite{wallace2021concealed}, and speech
recognition~\cite{zhai2021backdoor}; the acoustic vehicle-classification domain
has received comparatively little attention.

This paper makes the following contributions:
\begin{enumerate}[leftmargin=1.4em,itemsep=1pt,topsep=2pt]
\item We establish $p^* \leq 0.5\%$ as an empirical upper bound on the minimum
      poisoning rate for effective targeted attack on an acoustic vehicle
      classifier, corroborated across three independent seeds.
\item We prove analytically that aggregate accuracy monitoring is \emph{provably
      insufficient} for detecting targeted attacks against minority classes and
      characterize the bound as a function of class fraction.
\item We show that the spectrogram-domain backdoor trigger degenerates to a
      label-flip attack for minority classes---a previously undocumented interaction
      between attack type and dataset statistics.
\item We propose a verifiable pipeline architecture eliminating the trust
      assumptions exploited by both attack variants.
\end{enumerate}

\section{Related Work}

\textit{Data poisoning.} Biggio et al.~\cite{biggio2012poisoning} formalized
poisoning attacks as bilevel optimization. Subsequent work demonstrated
targeted clean-label attacks~\cite{shafahi2018poison} and gradient-matching
attacks~\cite{geiping2021witches}. Our setting requires labeled-sample
poisoning, which is simpler but sufficient for near-perfect targeted ASR.

\textit{Backdoor attacks.} BadNets~\cite{gu2019badnets} established the patch
backdoor framework in computer vision. Refinements include invisible
triggers~\cite{liu2018trojaning} and frequency-domain
perturbations~\cite{zeng2021rethinking}. For audio,
Zhai et al.~\cite{zhai2021backdoor} demonstrated ultrasonic backdoor triggers.
Our work is the first to characterize trigger-dominance collapse as a function
of class-minority ratio.

\textit{Acoustic vehicle classification.} MELAUDIS~\cite{melaudis2024} provides
labeled recordings from Melbourne urban intersections. Prior work focused on
feature engineering and model selection; security properties were not evaluated.

\textit{ML supply-chain security.} Work on ML-SBOM~\cite{bommasani2021},
dataset provenance~\cite{pushkarna2022datacard}, and cryptographic data
lineage~\cite{hutchinson2021c2pa} motivates pipeline-level defenses. We extend
these with Merkle commitments and post-quantum signatures.

\section{Dataset and Model}

\subsection{MELAUDIS Dataset}

MELAUDIS~\cite{melaudis2024} consists of stereo WAV recordings from fixed
microphone arrays at signalized intersections in Melbourne, Australia. After
filtering to single-vehicle events, the dataset comprises $\approx$9,600 clips
across six vehicle classes. Table~\ref{tab:dataset} shows the class distribution.

\begin{table}[h]
\caption{MELAUDIS class distribution (single-vehicle events).}
\label{tab:dataset}
\centering
\small
\begin{tabular}{lrr}
\toprule
\textbf{Class} & \textbf{Count} & \textbf{Fraction} \\
\midrule
Car        & $\approx$8,100 & 84.4\% \\
Tram       & $\approx$600   &  6.3\% \\
Truck      & $\approx$260   &  2.7\% \\
Bus        & $\approx$260   &  2.7\% \\
Motorcycle & $\approx$250   &  2.6\% \\
Bicycle    & $\approx$220   &  2.3\% \\
\bottomrule
\end{tabular}
\end{table}

The severe class imbalance---Car at 84\%, heavy vehicles at $\approx$3\%
each---is a structural property of real-world urban traffic directly exploited
by our attack.

\subsection{Features and Architecture}

Each clip is resampled to 16~kHz and converted to a 128-bin log-mel spectrogram
over 20--8000~Hz (25~ms window, 10~ms hop). We train a compact 2-D CNN
(four Conv-BN-ReLU-Pool blocks, global average pooling, 6-way softmax) using
Adam for 20 epochs with cross-entropy loss. The clean baseline achieves
\textbf{87.6\%} test accuracy.

\section{Attack Methodology}

\subsection{Threat Model}

We consider an attacker with \emph{write access to any unsigned stage of the
training pipeline}---annotation CSVs, feature manifests, split files, or
spectrogram generation code. This models a supply-chain adversary, compromised
annotation vendor, or malicious insider. The goal is to cause the model to
misclassify Truck as Car at inference while maintaining high aggregate accuracy
to evade detection.

\subsection{Attack 1: Targeted Label Flip}

At poisoning rate $p$, we uniformly sample $\lfloor p \cdot N \rfloor$ Truck
training records and relabel them as Car. Audio and spectrograms are unchanged.
The \emph{Attack Success Rate} is:
\[
  \mathrm{ASR} = \frac{\bigl|\{x \in \mathcal{D}_\text{test}^\text{Truck} : f_\theta(x) = \text{Car}\}\bigr|}{|\mathcal{D}_\text{test}^\text{Truck}|}
\]
We sweep $p \in \{0.5\%, 1\%, 2\%\}$ with three seeds per rate.

\subsection{Attack 2: Backdoor Patch}

The backdoor variant extends Attack~1 by additionally stamping a $12{\times}12$-pixel
bright white square in the bottom-right corner of each poisoned Truck spectrogram.
This spatial region corresponds to a brief, high-frequency artifact at the end of
the clip that is inaudible in practice (near-zero RMS difference from clean audio).
We evaluate:
\begin{itemize}[leftmargin=1.4em,itemsep=0pt,topsep=2pt]
  \item \textbf{Clean ASR}: fraction of unpatched Truck test samples predicted
        as Car.
  \item \textbf{Triggered ASR}: fraction of patched Truck test samples predicted
        as Car.
\end{itemize}

\section{Results}
\label{sec:results}

\subsection{Targeted Label Flip}

Table~\ref{tab:label_flip} summarizes results across poisoning rates.

\begin{table}[h]
\caption{Targeted label-flip: Truck $\to$ Car.}
\label{tab:label_flip}
\centering
\small
\begin{tabular}{cccc}
\toprule
\textbf{Rate} & \textbf{Flipped} & \textbf{Accuracy} & \textbf{ASR (95\% CI)} \\
\midrule
0 (clean)& 0   & 87.6\% & $\approx$0\% \\
0.5\%    & 48  & 87.6\% & \textbf{95.7\%} [88.4, 100] \\
1.0\%    & 96  & 87.4\% & 94.9\% [83.8, 100] \\
2.0\%    & 192 & 86.4\% & 95.7\% [92.0, 99.4] \\
\bottomrule
\end{tabular}
\end{table}

\textbf{Finding 1 ($p^* \leq 0.5\%$).} Corrupting 48 training labels achieves
95.7\% Truck$\to$Car ASR. The true minimum may be even lower; 0.5\% was already
fully effective.

\textbf{Finding 2 (Complete stealth).} Across all rates, overall test accuracy
is statistically indistinguishable from baseline. Standard evaluation metrics
observe \emph{no anomaly}.

\textbf{Finding 3 (Class imbalance amplifies stealth).} Let $\beta$ be the
fraction of training examples in the targeted minority class. A complete
collapse of that class contributes at most $\beta$ to aggregate accuracy loss.
For Truck ($\beta \approx 3\%$), a 96\% ASR reduction in Truck accuracy moves
aggregate accuracy by $3\% \times 96\% \approx 2.9\%$---below training-run
variance. Formally:
\[
  \Delta\text{Acc}_\text{max} \leq \beta
\]
This bound holds regardless of architecture or attack method. The more
safety-critical and rare the target class, the more invisible the attack.

\subsection{Backdoor Patch}

Table~\ref{tab:backdoor} shows results for $p = 0.5\%$.

\begin{table}[h]
\caption{Backdoor patch attack ($p{=}0.5\%$, Truck $\to$ Car).}
\label{tab:backdoor}
\centering
\small
\begin{tabular}{lc}
\toprule
\textbf{Metric} & \textbf{Value} \\
\midrule
Overall test accuracy       & 87.44\%  \\
Clean ASR (unpatched Truck) & 94.87\%  \\
Triggered ASR (patched Truck) & 94.87\% \\
\bottomrule
\end{tabular}
\end{table}

\textbf{Finding 4 (Trigger-dominance collapse).} Clean ASR equals Triggered ASR,
and both are high---the patch trigger is functionally redundant. At $p = 0.5\%$
on a class with $\approx$182 training samples, we poison $\approx$26\% of Truck
examples. With so few clean Truck samples remaining, the model cannot form a
coherent Truck representation regardless of whether the patch appears at inference.

\textbf{Implication.} For minority classes in heavily imbalanced datasets, the
backdoor attack degenerates to a label-flip attack. The transition occurs at
poisoning rates that are operationally trivial. An attacker need not modify
spectrograms; labels alone are the vulnerability. The patch adds injection-vector
stealth (harder to detect than a label diff) but provides no additional attack
capability.

\section{Attack Surface Analysis}

Table~\ref{tab:surface} catalogs unsigned pipeline stages and their exposure.

\begin{table}[h]
\caption{Training pipeline attack surface.}
\label{tab:surface}
\centering
\small
\begin{tabular}{lcc}
\toprule
\textbf{Stage} & \textbf{Label Flip} & \textbf{Backdoor} \\
\midrule
Raw WAV files             & Low   & \textbf{High} \\
Annotation CSVs           & \textbf{High}  & Low  \\
Feature manifest          & \textbf{High}  & \textbf{High} \\
Spectrogram generation    & Low   & \textbf{High} \\
Train/val/test splits     & \textbf{High}  & Medium \\
Model checkpoint          & N/A   & \textbf{High} \\
\bottomrule
\end{tabular}
\end{table}

Table~\ref{tab:controls} maps defensive controls to detection capability.
\emph{No single control catches both attacks.} Critically, aggregate accuracy
monitoring---the most commonly deployed safeguard---catches neither.

\begin{table}[h]
\caption{Detection controls by attack type.}
\label{tab:controls}
\centering
\small
\begin{tabular}{lcc}
\toprule
\textbf{Control} & \textbf{Label Flip} & \textbf{Backdoor} \\
\midrule
Overall accuracy monitoring        & No            & No \\
Per-class F1 / confusion matrix    & \textbf{Yes}  & No \\
Hash on label manifests            & \textbf{Yes}  & Partial \\
Hash on spectrogram files          & \textbf{Yes}  & \textbf{Yes} \\
Signed pipeline artifacts          & \textbf{Yes}  & \textbf{Yes} \\
Feature distribution anomaly det.  & Partial       & \textbf{Yes} \\
Spectrogram visual audit           & Partial       & \textbf{Yes} \\
\bottomrule
\end{tabular}
\end{table}

\section{Trust-Minimized Pipeline Defense}

\subsection{Architecture Overview}

We propose a cryptographically verifiable training pipeline that eliminates
the trust assumptions exploited by both attacks. Each stage signs its output
before the next stage consumes it; any mutation invalidates the signature chain.

\begin{figure}[h]
\centering
\begin{tikzpicture}[
  node distance=0.4cm and 0.2cm,
  box/.style={draw, rounded corners=2pt, font=\scriptsize\ttfamily,
              inner sep=3pt, minimum width=1cm, minimum height=0.45cm},
  arrow/.style={-{Stealth[length=3pt]}, thin},
  sig/.style={font=\scriptsize, color=green!50!black, above=-1pt}
]
\node[box] (wav)  {WAV};
\node[box, right=0.6cm of wav] (csv)  {CSV};
\node[box, right=0.6cm of csv] (feat) {Feat.};
\node[box, right=0.6cm of feat] (spl)  {Splits};
\node[box, right=0.6cm of spl]  (mdl)  {Model};

\draw[arrow] (wav)  -- node[sig]{$\sigma_d$} (csv);
\draw[arrow] (csv)  -- node[sig]{$\sigma_a$} (feat);
\draw[arrow] (feat) -- node[sig]{$\sigma_p$} (spl);
\draw[arrow] (spl)  -- node[sig]{$\sigma_o$} (mdl);

\node[font=\tiny, below=0.1cm of wav]  {device};
\node[font=\tiny, below=0.1cm of csv]  {annotator};
\node[font=\tiny, below=0.1cm of feat] {pipeline};
\node[font=\tiny, below=0.1cm of spl]  {orchestrator};
\end{tikzpicture}
\caption{Trust-minimized pipeline. Each stage signs its output ($\sigma_i$)
before the downstream stage consumes it.}
\label{fig:pipeline}
\end{figure}
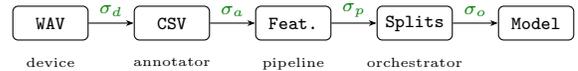

\subsection{Merkle Dataset Commitment}

After feature generation, we commit the full dataset to a Merkle tree. Each
leaf hashes the sample identifier, label, and file hashes:
\[
  \ell_i = \mathrm{SHA}\text{-}256(\,\mathtt{id}_i \;\|\; y_i \;\|\; h^\text{wav}_i \;\|\; h^\text{feat}_i\,)
\]
The Merkle root $r$ commits to all $N$ samples. Any single label flip changes
a leaf, propagates up the tree, and changes $r$, invalidating the signature.

Key properties: (1) inclusion proofs require $O(\log N)$ hashes for per-sample
verification; (2) the root may be published to an append-only log (git,
Certificate Transparency, or on-chain), preventing retroactive modification;
(3) the trainer verifies $r$ before loading any data and aborts on mismatch.

We implement this in \texttt{src/melaudis/merkle.py}. A poisoned manifest
produces the terminal output:
\begin{center}
\small\ttfamily
\textcolor{red}{$\times$ MERKLE ROOT MISMATCH --- ABORT.}
\end{center}

\subsection{Post-Quantum Signatures}

Classical signature schemes (Ed25519, ECDSA) are broken by Shor's
algorithm~\cite{shor1994algorithms}. For ML training data---which may anchor
model lineage for a decade or more---this creates a \emph{harvest-now,
forge-later} threat: an adversary archives signed manifests today and forges
poisoned provenance records once a quantum computer is available.

We use \textbf{ML-DSA-65} (CRYSTALS-Dilithium3, NIST FIPS~204,
2024)~\cite{nist2024fips204} for all pipeline manifest signatures.
Security rests on module lattice problems, unaffected by Shor's or Grover's
algorithms. SHA-256 in the Merkle tree retains 128-bit post-quantum security
under Grover's algorithm and requires no modification.

\subsection{Tiered Defense Summary}

\textbf{Tier 1 (Cryptographic, highest ROI):}
SHA-256 content-addressed artifacts at every stage; ML-DSA-65 signed manifests
verified before each downstream stage; append-only artifact storage.

\textbf{Tier 2 (Statistical monitoring):}
Per-class F1 and confusion matrix alerts after each training run; spectrogram
intensity distribution anomaly detection; label distribution drift alerts.

\textbf{Tier 3 (Architectural hardening):}
ML-SBOM with certified data provenance; differential training audits (shadow
models on data subsets); certified robustness via randomized smoothing;
inference-time trigger detection on input spectrograms.

\section{Discussion}

\subsection{Generalization Across Sensing Modalities}

The two core mechanisms demonstrated here---class-imbalance-amplified stealth
and minority-class trigger-dominance collapse---are modality-agnostic. The
bound $\Delta\mathrm{Acc}_\mathrm{max} \leq \beta$ holds for any classification
task where the targeted class comprises fraction $\beta$ of the training corpus,
regardless of sensor type, feature representation, or model architecture.
We enumerate four high-consequence sensing domains where the same structural
vulnerability applies directly.

\textit{Thermal / FLIR imaging.} Forward-looking infrared classifiers used in
perimeter security, autonomous vehicles, and aerial surveillance routinely
operate on heavily imbalanced datasets: pedestrians and cyclists are rare
relative to background and vehicle classes, particularly at night or in
low-traffic environments. A targeted poisoning attack relabeling a small
fraction of pedestrian training frames as background would suppress pedestrian
detection while leaving aggregate classification accuracy essentially unchanged.
The spectrogram-domain backdoor analogue is a localized thermal patch---a
pixel-region anomaly invisible to the human eye under normal operating
conditions but detectable only by feature-level inspection.

\textit{RF spectrum sensing.} Wideband spectrum classifiers for cognitive radio,
drone detection, and electronic warfare must identify rare emitter types
(specific drone RF signatures, low-probability-of-intercept waveforms) against
a dominant background of known signals. Minority emitter classes can comprise
$<$1\% of training captures in realistic deployment data. The $\beta$ bound
implies that complete suppression of a rare emitter class is undetectable
via aggregate classification accuracy, which is the primary evaluation metric
in most spectrum sensing benchmarks. A backdoor trigger in the frequency
domain---a narrowband injection at a fixed offset---is the RF analogue of
the spectrogram patch and similarly degenerates to a label-flip attack for
rare emitter classes.

\textit{Radar (automotive and air traffic).} Radar-based classifiers for
low-radar-cross-section (RCS) targets---cyclists and pedestrians in automotive
radar, small UAVs in air traffic management---face the same imbalance structure.
Poisoning a small fraction of low-RCS target labels as ``clutter'' or
``background'' suppresses detection of the most safety-critical targets
(small and slow-moving objects) while leaving bulk detection metrics for
large, easy targets (trucks, aircraft) unaffected.

\textit{LiDAR point-cloud classification.} Autonomous vehicle perception
systems trained on LiDAR point clouds classify sparse classes (cyclists,
debris, animals) against a dominant majority of road surface, vegetation,
and passenger vehicles. The trigger-dominance collapse observed here
predicts that backdoor attacks targeting sparse point-cloud classes will
degenerate to label-flip attacks, since the poisoned fraction rapidly
exhausts the available training examples for that class. This has direct
implications for the security of sensor-fusion stacks, where a compromised
LiDAR classifier feeds downstream planning modules.

In each domain, the attack surface is identical to the one formalized in
Section~\ref{sec:surface}: any unsigned stage between raw sensor capture
and model training is a viable injection point, and the defense architecture
proposed in Section~VII applies without modification.

\subsection{Responsible Disclosure}

The dataset, architecture, and poisoning methodology are publicly available
or straightforward to replicate from public components. We do not disclose
novel zero-day attack primitives; our contribution is quantifying known attack
types on an underexplored domain and proposing concrete mitigations. Full
pipeline code, including Merkle verification tooling, is available in the
companion repository.

\subsection{Limitations}

(1)~We evaluate a single compact architecture; ASR may differ for larger or
more regularized models. (2)~Truck has only $\approx$182 training samples; high
ASR at $p = 0.5\%$ partially reflects the small absolute class size. (3)~The
backdoor trigger is a fixed visible patch; invisible or feature-space triggers
may exhibit different trigger-dominance dynamics.

\section{Conclusion}

We have shown that an acoustic vehicle classifier can be induced to misclassify
95.7\% of trucks as cars by corrupting 48 training labels (0.5\% of the
dataset), with zero detectable change in aggregate accuracy. This stealth is
structural: the maximum detectable accuracy drop from a complete targeted
attack is bounded by the minority class fraction $\beta$. For the typical
class imbalances found in real-world sensing deployments, this bound falls
below training-run noise.

The backdoor variant reveals a further structural result: when the targeted
class is a small minority, the patch trigger degenerates to a label-flip attack.
This simplifies the adversary's task and shifts the relevant attack surface
entirely to annotation labels.

Against both attacks, the only reliable defenses are those that eliminate trust
in unsigned pipeline artifacts: cryptographic artifact hashing, Merkle-tree
dataset commitment, and post-quantum pipeline signatures. Aggregate accuracy
monitoring---the most common deployed safeguard---is provably insufficient.

\section*{Acknowledgments}

The authors thank the creators of the MELAUDIS dataset for making real-world
urban intersection recordings publicly available for research.


\end{document}